\documentclass[
              ,aps
              ,prl
              ,floatfix
	           ,footinbib
              ,a4paper
              ,preprint    
              ,endfloats   
              ]{revtex4}

\usepackage{longtable}
\usepackage{rotating}
\usepackage{amsmath}
\usepackage{textcomp}
\usepackage{graphics}

\newif\ifpdf\ifx\pdfoutput\undefined\pdffalse\else\pdfoutput=1\pdftrue\fi
\newcommand{\pdfgraphics}{\ifpdf\DeclareGraphicsExtensions{.pdf,.jpg}\fi}

\begin{document}
\pdfgraphics

\title{
Origin of the anomalous piezoelectric response in wurtzite Sc$_x$Al$_{1-x}$N alloys
}

\author{Ferenc Tasn\'adi}
\email{tasnadi@ifm.liu.se}
\author{Bj\"orn Alling}
\author{Carina H\"oglund}
\author{Gunilla Wingqvist}
\author{Jens Birch}
\author{Lars Hultman}
\author{Igor A. Abrikosov}
\affiliation{Department of Physics, Chemistry and Biology
(IFM), Link\"oping University, SE-581 83 Link\"oping, Sweden}

\date{\today}
%
\begin{abstract}
The origin of the anomalous, 400\% increase of the piezoelectric coefficient in Sc$_x$Al$_{1-x}$N
alloys is revealed. Quantum mechanical calculations show that the effect is intrinsic. It comes
from a strong change in the response of the internal atomic coordinates to strain 
and pronounced softening of C$_{33}$ elastic constant.
The underlying mechanism is the flattening of the energy landscape due to a
competition between the parent wurtzite and the so far experimentally unknown hexagonal phases of
the alloy. Our observation provides a route for the design of materials with high piezoelectric
response.
\end{abstract}
\maketitle

Piezoelectricity is a phenomenon whereby the material becomes electrically
polarized upon the application of stress. Discovered in 1880 by Pierre and
Jacques Curie, the piezoelectric effect is used in many modern devices, like
touch-sensitive buttons, cell phones, computers, GPS, bluetooth, WLAN etc.
Moreover, new piezoelectric materials where excellent piezoelectric response
is combined with high temperature operation capabilities are currently
sought. A fundamental understanding of factors, which influence the
piezoelectric properties of materials appears therefore to be of high
scientific and technological values.

Recently Akiyama {\it et al.}\,\cite{piezo_ScAlN} have discovered a tremendous $\sim$400\% increase
of the piezoelectric moduli d$_{33}$ in Sc$_x$Al$_{1-x}$N alloys in reference to pure wurtzite AlN
around x = 0.5. This is the largest piezoelectric response among the known tetrahedrally boundend semiconductors.
Since AlN can be used as a piezoelectric material at temperatures up to 1150\,\textcelsius\ and it easily can be
grown as $c$-oriented, the AlN-based alloys with such a high response open a route for dramatic
increase in overall performance of piezoelectric based devices. Nevertheless, a
fundamental understanding of the phenomenon leading to such a dramatic improvement in piezoelectric
properties of AlN is absent. In particular, it is unclear if the enhanced piezoelectric response in
Sc$_x$Al$_{1-x}$N is related to the microstructure or it is an intrinsic effect of the alloying.
The aim of
this Letter is to present the solutions to this problem from ab-initio theory.

Reliable {\it ab-initio} calculation of polarization in solids have became possible only with
the formulation of the Berry-phase approach by King-Smith and Vanderbilt\,\cite{King-Smith_Vanderbilt,RMP_Resta} in mid-90s.
However, though alloying is known as one of the most efficient ways of improving materials performance,
first-principles studies of piezoelectric response of disordered alloys have appeared only recently
\,\cite{AvdW_tensorial_01,tasnadi_wBAN}.
In this Letter we show that alloy physics, associated with the tendency towards
different coordination preference in different local chemical environments, can lead to a strong change in
polarization induced by internal distortions and pronounced softening of C$_{33}$ elastic constant,
giving rise to the large piezoelectric effect in Sc$_x$Al$_{1-x}$N.

The calculations of piezoelectric coefficients in wurtzite Sc$_x$Al$_{1-x}$N alloys performed
in this study are based on density-functional theory within the generalized gradient
approximation\,\cite{PBE_made_simple} either by using the projector augmented wave\,\cite{PAW}
technique implemented into VASP\,\cite{VASP} or the Vanderbilt ultrasoft-pseudopotential scheme\,\cite{ultrasoft_PP} within
the Quantum-ESPRESSO program package\,\cite{QE-2009}. The proper piezoelectric tensor components $e_{33}$
were calculated with the modern Berry-phase
approach\,\cite{King-Smith_Vanderbilt} as proposed by Vanderbilt\,\cite{Vanderbilt_JPCS}.
To model the alloys, we applied the special quasirandom structure (SQS) method\,\cite{Zunger_SQS},
where the ideally random alloys are modeled with cleverly designed ordered super-structures. The
128-atoms wurtzite based SQS supercells were generated for x=0.125, 0.25,
0.375, and 0.50 by optimizing the Warren-Cowley pair short-range order parameters\,\cite{RPP_Ruban_Abrikosov} up to the 7th shell.
The Sc atoms were substituted only on the Al sublattice.

\begin{figure}[h!]
\includegraphics{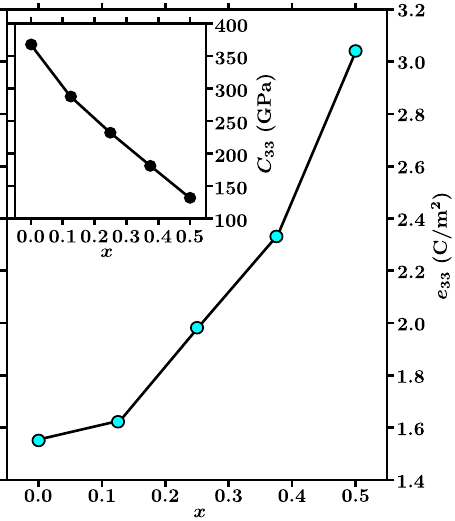}
\caption{\label{fig_piezo}(Color online)
The piezoelectric $e_{33}$ and elastic stiffness $C_{33}$ constants of wurtzite
Sc$_x$Al$_{1-x}$N alloys.}
\end{figure}

Figure\,\ref{fig_piezo} demonstrates the resulting behavior of the piezoelectric constant
$e_{33}$ of Sc$_x$Al$_{1-x}$N versus composition. $e_{33}$ shows a smooth, but non-linear
increase with the amount of Sc, reaching a large, factor of two increment at $x=0.5$.
The inset of this figure displays the calculated stiffness constants $C_{33}$, which
exhibit an almost linear decrease. By taking
the roughly two-fold decrease of $C_{33}$ and the upper bound of
$C_{13}\approx C_{13}^{\mathrm{AlN}}=98$ GPa\,\cite{tasnadi_wBAN}, the acceptable general rule of
$R=-\left(d_{33}/d_{31}\right)\geq 2$\,\cite{d33_definition} ensures that $RC_{33}\gg C_{13}$. 
Consequently, we can reproduce the approximately 400\% increase of the piezoelectric moduli 
$d_{33}\approx \left(e_{33}/C_{33}\right)$ observed  experimentally around $x=0.5$.
These results clearly establish that the observed increase of $d_{33}$ (with units C/N) is
definitely inherent for Sc$_{x}$Al$_{1-x}$N alloys. Hence, our random alloy model
should provide valuable insight to explain the microscopic physical origin of the highly
increased piezoelectric response in these alloys. The quantitative agreement between
our ab-initio, 0~K results with the reported room temperature experimental results\,\cite{piezo_ScAlN}
allows us to expect that the phonon contribution to piezoelectricity in ScAlN alloys is small 
at low and moderate temperatures.

The large non-linearity of $e_{33}$ higly diminish the impact of the linear, compositional
alloying and points to other alloying related effects, such as internal strain\,\cite{nonlinear_P33}.
Consequently, the physical origin of the large piezoelecetric
response should be requested more in relation to intrinsic structural properties of these alloys.
Though, in the
{\it ab-initio} computational technique we applied supercell geometries, the further analysis of the
resulting data is done by projecting to a single wurtzite unit cell using site-averaging.
Therefore, the piezoelectricity can be discussed like for III-V nitrides\,\cite{single_wurtzite}
by using the expression,
\begin{equation}
e_{33}(x)=e_{33}^{\text{\it clamped-ion}}(x)+
\frac{4\,e\,Z^{\ast}(x)}{\sqrt{3}\,a(x)^2}\,\frac{du(x)}{d\delta},
\label{eq_piezo}
\end{equation} where $e$ is the electronic charge, $a$ stands for the equilibrium lattice
parameter, $u$ for the wurtzite internal parameter and $Z^{\ast}$ is the dynamical Born
or transverse charge in units of $e$. $\delta$ is the macroscopic applied strain.
The first, {\it clamped-ion} term expresses the electronic response to strain, while the second
term describes the effect of internal strain on the piezoelectric polarization.

\begin{figure}[h!]
\includegraphics{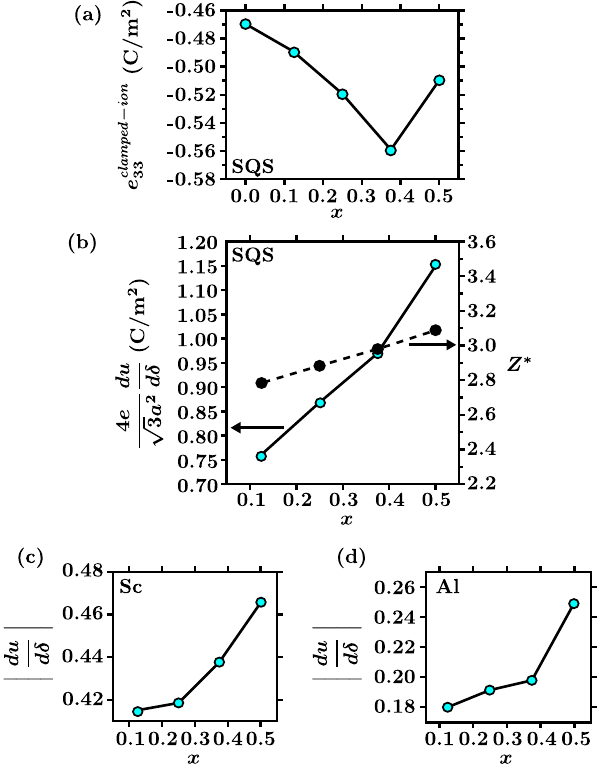}
\caption{\label{fig_du_ddelta}(Color online)
The calculated contributions of $e_{33}$ as introduced by Eq.(\ref{eq_piezo}). (a) shows
the {\it clamped-ion} term. (b) gives the strain contributions and the Born charges $Z^{\ast}$
calculated from Eq.(\ref{eq_piezo}). (c) and (d) exhibit the Sc and Al site resolved
internal strain sensitivity of wurtzite Sc$_x$Al$_{1-x}$N.}
\end{figure}

Although the {\it clamped-ion} term is not expected to contribute to the
strong increase of $e_{33}$, it was considered in our investigation as the first term in
Eq.(\ref{eq_piezo}). As Fig.\,\ref{fig_du_ddelta} (a) shows, a monotonous decrease
(from -0.47 to -0.56 C/m$^2$) has been found for this quantity up to $x=0.375$,
followed by an increase up to -0.51 C/m$^2$ at $x=0.5$. The more important characteristic of the
local structural sensitivity to macroscopic axial strain $\delta$ is expressed in the second
term of Eq.(\ref{eq_piezo}).
Figures\,\ref{fig_du_ddelta}(b)-(d) show our calculated compositional
weighted and site resolved contributions. The figures unambiguously demonstrate, that 
the change in polarization in these alloys is mainly induced by internal distorsions
around the Sc sites. One finds a factor of two higher sensitivity of $u$-parameter
for the Sc sites than for the Al ones. At low Sc fraction the value of
$du/d\delta$ approaches $-0.18$, which is just the value of pure
wurtzite AlN. This justifies our projected single cell approach.

The kink observed for
the {\it clamped-ion} term at $x=0.375$ in Fig.\,\ref{fig_du_ddelta}(a)
seems to be connected to those shown in in Fig.\,\ref{fig_du_ddelta}(b) and (d). Hence, this kink
refers to the rapidly increasing internal strain sensitivity around the Al sites, which should be
determined by electronic effects. To adequately discuss the strong increase of
$e_{33}$ in Fig.\,\ref{fig_piezo}, based on the terms introduced in Eq.(\ref{eq_piezo}), one should consider
also the dielectric effects represented by the dynamical Born charge $Z^{\ast}$.
The dashed line in Fig.\,\ref{fig_du_ddelta}(b) shows $Z^{\ast}$ calculated from Eq.(\ref{eq_piezo}).
$Z^{\ast}$ varies within about 15\% around the expected ionic nominal value of $3$.
In comparison, the structural strain part shown by solid line in Fig.\,\ref{fig_du_ddelta}(b) changes
by around 60\%, which derogates the dielectric
contribution associated with $Z^{\ast}$. Consequently, the strongly increased internal
axial strain sensitivity, coming primarily from the Sc sites, is mostly responsible for
the non-linear enhancement of the piezoelectric constant $e_{33}$ in wurtzite Sc$_x$Al$_{1-x}$N
alloys. Therefore in this system the observed large piezoelectric response and large elastic
softening should be governed by local structural instabilities.

\begin{figure}[h!]
\includegraphics[width=5.4cm]{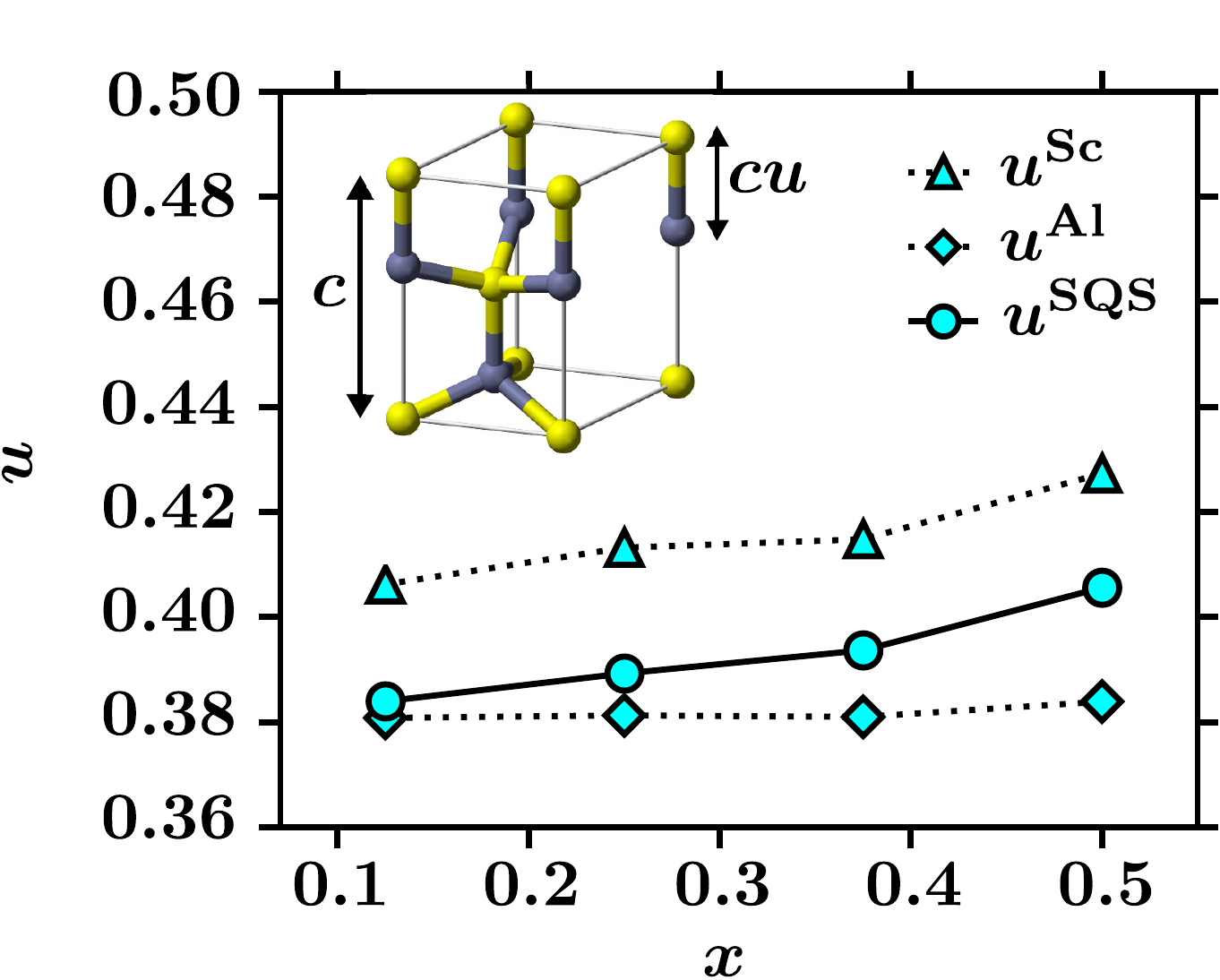}
\caption{\label{fig_average_u}(Color online)
The averaged wurtzite internal parameter $u$ as a function of Sc concentration. The dotted lines
show the Sc and Al site resolved values, while the solid line is the compositional weighted of
Sc$_x$Al$_{1-x}$N.}
\end{figure}

From a structural point of view, the parent wurtzite AlN (w-AlN)
has the pyroelectric point group symmetry $6mm$ and the value $u=0.38$ of the wurtzite internal
parameter. The inset of Fig.\,\ref{fig_average_u} exposes the wurtzite (B4) structure with its
four-fold (tetrahedral) coordination and $u$ parameter. Thus, w-AlN exhibits spontaneous polarization
and three non-vanishing independent piezoelectric tensor components. The ground state crystal structure
of the other parent ScN is cubic rock-salt, but the existence of a metastable nearly five-fold
coordinated layered hexagonal phase (h-ScN) with a flattened bilayer structure $u=0.5$
(see the inset of Fig.\,\ref{fig_average_u}) has been proposed theoretically\,\cite{hex_ScN}. Its wurtzite phase
turned out to be unstable. Accordingly, h-ScN shows the centrosymmetric point group $6/mmm$
and thus poses no polar properties.

Fig.\,\ref{fig_average_u} shows the calculated wurtzite internal parameter $u$ of Sc$_x$Al$_{1-x}$N as a
function of Sc concentration. Interestingly, around the Al sites,
the found average $u^{\mathrm{Al}}$ value shows strong stability, it is actually frozen to its parent w-AlN
value, $u=0.38$.  Around the Sc atoms one finds a monotonous increase of $u^{\mathrm{Sc}}$,
though it is still zwell below the value 0.5 of the hexagonal phase. In accordance with Fig.\,\ref{fig_du_ddelta}(c) and
(d), this behavior is in clear correspondence with the larger sensitivity of the Sc sites to axial strain.

The possible phase transition from wurtzite to hexagonal structure, set up by the parent
binary phases, can in principle result in enhanced piezoelectric behavior. This issue has been
discussed in detail for ordered Sc$_x$Ga(In)$_{1-x}$N  alloys\,\cite{ScGaN_ScInN}.
Nonetheless the direct influence of this phase transition on our results, in general,
can be ruled out. One would observe a jump in the structural internal parameter
$u$\,\cite{ScN_strain}. Our calculated composition weighted average wurtzite internal parameters
$u$, as shown in Fig.\,\ref{fig_average_u}, display instead a small and smooth increase
from 0.384 to 0.406. Accordingly, the N coordination in these alloys is still fairly tetrahedral
and not five-fold. In the nearly five-fold coordinated hexagonal phase at $x=0.5$ we found the value
$u=0.49$ corresponding to insignificant piezoelectric activity. The tendency toward different coordination
preference in different local chemical enviroments is evident from our calculations.

\begin{figure}[h!]
\includegraphics[width=7.9cm]{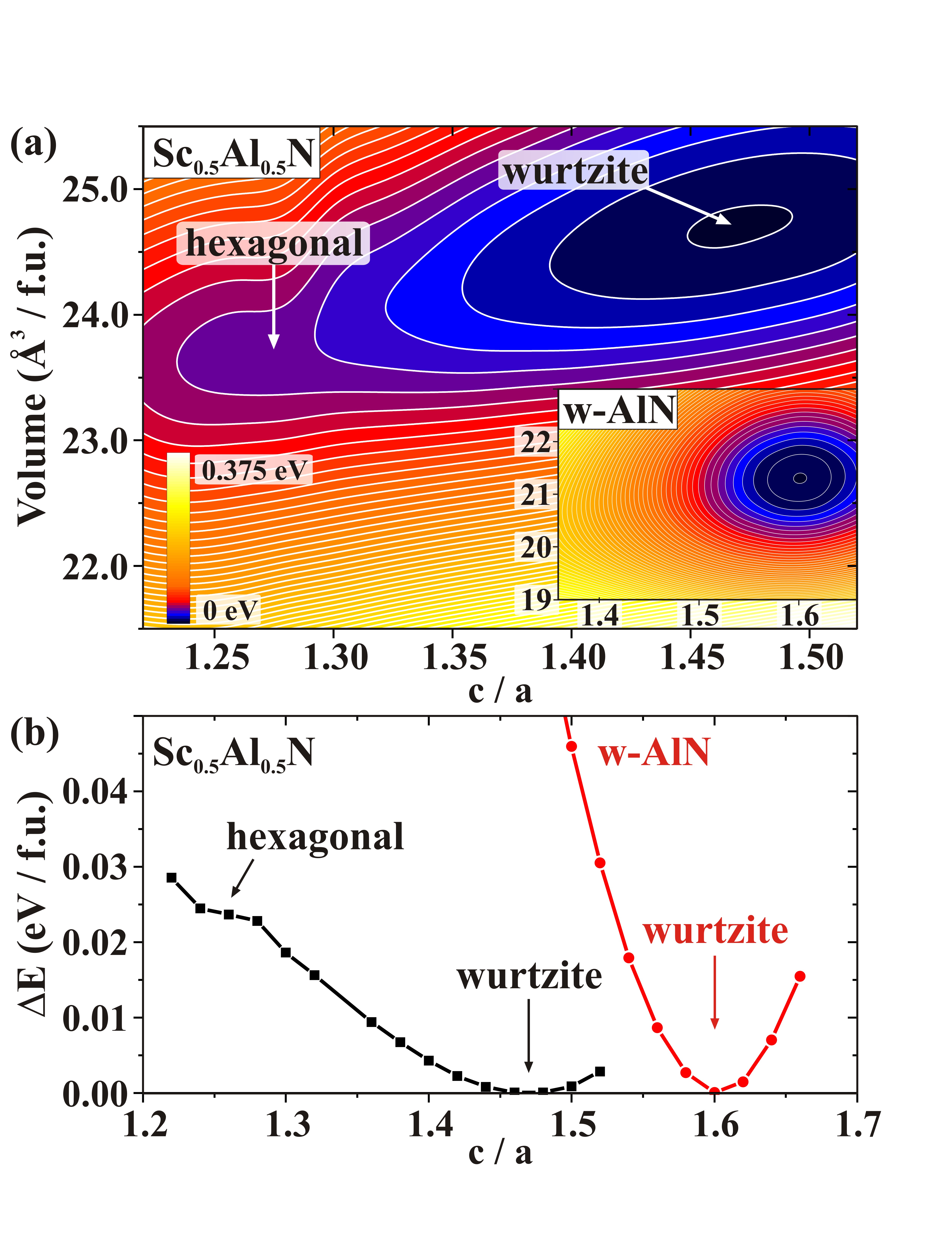}
\caption{\label{fig_energy_landscape}(Color online)
The energy landscape of wurtzite Sc$_{0.5}$Al$_{0.5}$N. (a) presents it as a function of $c/a$
and volume with both, wurtzite and hexagonal phases of the alloy, while the inset shows the
energy surface of the parent wurtzite AlN. (b) shows minimum energy cross sections of the
energy landscapes.}
\end{figure}

In order to gain microscopic insights into the observed huge axial softening responsible
for the anomalous increase of the piezoelectric coefficient in Sc-doped AlN, as indicated
by Fig.\,\ref{fig_piezo}, we calculated the energy
landscape of Sc$_{0.50}$Al$_{0.50}$N showing the largest enhancement of $e_{33}$.
The resulting landscape is presented in Fig.\,\ref{fig_energy_landscape}(a)
where each contour line represents an energy step of 5 meV/f.u. 
The landscape of pure w-AlN is
shown in the inset of Fig.\,\ref{fig_energy_landscape}(a) for comparison and reflects how
strongly the elastic properties are affected by Sc alloying. The shallow region in (a)
connects the wurtzite derived global energy minimum with a residue of the h-ScN phase present
at $c/a\approx1.27$. The topology of the minimum energy region is considerably alongated in
the $c/a$ direction, which guarantees that it
affects the alloy properties mostly along lattice parameter $c$. Since the elastic
stiffness constant $C_{33}$ is the curvature of the total energy with respect to $c$, it gives
an explanation of our results shown in the inset of Fig.\,\ref{fig_piezo}.
Note that the existence of a Gibbs free-energy flattening in ferroelectric perovskite crystals
has been discussed by Budimir~\emph{et al.}\,\cite{Damjanovic_flattening} in the framework
of the phenomenological Landau-Ginsburg-Devonshire theory. Here, it appears explicitely as a result
of {\it ab-initio} quantum mechanical calculations.

To interpret these results one can consider the different coordination preferences
of ScN and AlN. The tetrahedral coordination of nitrogen in w-AlN is strikingly different as
compared to the octahedral coordination in rock salt ScN and also to the nearly five-fold
coordination in h-ScN. In Sc$_x$Al$_{1-x}$N-alloys the Al and Sc atoms compete
about the coordination of nitrogen thus weakening both the internal resistance against
nitrogen displacement as well as the resistance against changing the $c/a$-ratio. As the
Sc composition increases this competition becomes more evident\,\cite{Hoglund_JAP_xxx} and
at $x=0.50$ the structural parameters have become extremely sensitive to external stress underlined
by the near energy degeneracy of
the two parent phases. The previously discussed kink in the piezoelectric
{\it clamped-ion} term and in the strain sensitivity shown in Fig.\,\ref{fig_du_ddelta}(d)
clearly demonstrate the amplified effect. The fact that the h-ScN derived phase of Sc$_{0.50}$Al$_{0.50}$N is not
really a local energy minimum with respect to change in $c/a$, but rather an energy saddle
point is seen clearly in Fig.\,\ref{fig_energy_landscape}(b). The absence of an
energy barrier between the two phases is likely to be connected to the fact that we are considering a real random
alloy where the presence of a chemically different local environments around the nitrogen atoms, all
with different preferences for the local coordination, smears out the dependence of the energy on the
$c/a$-ratio. This smearing results in smaller curvature, what is seen as a decrease of
the elastic constant $C_{33}$ (inset of Fig.\,\ref{fig_piezo}). Thus, it is underlined that
the high piezo-electric response in the Sc$_x$Al$_{1-x}$N system is really due to alloy
physics of the wurtzite phase rather than an actual phase transition. Nevertheless, the
close presence, both in energy and volume, of the similar phases with different
coordination preference, as in Sc$_{0.50}$Al$_{0.50}$N, seems to be the signature of an
ideal semiconductor alloy system with a potential for large piezoelectric responses.
It can thus be used as a criteria in search of better piezo-electric materials.

In summary, we present the theory that reveals the origin of the observed anomalous enhancement
of piezoelectric response in wurtzite Sc$_x$Al$_{1-x}$N alloys. Our first-principles calculations
confirm that the 400\% increase of the piezoelectric constant is an intrinsic alloying effect.
The energy-surface topology is found to be strongly influenced by the alloying, being elongated around
the global minimum along $c/a$ direction. This leads to the large elastic softening along the crystal
parameter $c$, and raises significantly the intrinsic sensitivity to axial strain resulting in the
highly increased piezoelectric constant. The role of local environment effects characteristic for disordered
alloys is demonstrated. The preference of Sc atoms to bind rather hexahedrally than tetrahedrally to
nitrogen results in a frustrated system with a strong response to strain. The effect is particularly
accentuated at intermediate compositions where the elongated double-minimum energy landscape is flattened
due to the energy proximity of the wurtzite and hexagonal phases of these alloys.
Note also, that the hexagonal phase of Sc$_x$Al$_{1-x}$N alloy at 50\% composition appears at the saddle point
rather than in an energy minimum. Therefore, it is dynamically unstable, and most probably could not be
synthesized experimentally.
Accordingly, we conclude as a rule, that structural phase competition similar to the one observed in this work
can be a general key-point to search for materials systems with substantial enhancement of the piezoelectric
response. This establishes a novel route for the design of materials to be used in piezoelectric applications.

The authors thank P. Muralt for a useful discussion.
The Swedish Foundation for Strategic Research (SSF) via the
MS$^2$E Research Center, The G\"oran Gustafsson Foundation for Research
in Natural Sciences and Medicine, the MultiFilms
SSF project and The Swedish Research Concil (VR) are acknowledged for financial support. 
Calculations have been performed at the Swedish National Infrastructure for Computing (SNIC). 
\bibliographystyle{unsrt}

\end{document}